\documentclass[aps,pra,showpacs,twocolumn,groupedaddress]{revtex4}
\usepackage{amsmath}
\usepackage{amssymb}
\usepackage{graphics}
\usepackage[dvips]{graphicx}
\usepackage{graphicx}
\usepackage{color}

\newcommand \beq{\begin{eqnarray}}
\newcommand \eeq{\end{eqnarray}}
\def\simge{\mathrel{%
       \rlap{\raise 0.511ex \hbox{$>$}}{\lower 0.511ex \hbox{$\sim$}}}}
\def\simle{\mathrel{
       \rlap{\raise 0.511ex \hbox{$<$}}{\lower 0.511ex \hbox{$\sim$}}}}

\begin{document}
\title{Transport in very dilute solutions of $^3$He in superfluid $^4 $He}
\author{Gordon Baym,$^{a,b}$ D.\ H.\ Beck,$^a$ and C.\ J.\  Pethick$^{a,b,c}$}
\affiliation{\mbox{$^a$Department of Physics, University of Illinois, 1110
  W. Green Street, Urbana, IL 61801} \\
\mbox{$^b$The Niels Bohr International Academy, The Niels Bohr Institute,}\\
\mbox{Blegdamsvej 17, DK-2100 Copenhagen \O, Denmark}\\	
\mbox{$^c$NORDITA, KTH Royal Institute of Technology and Stockholm University, Roslagstullsbacken 23, SE-10691 Stockholm, Sweden} 
}

\date{\today}

\begin{abstract}

Motivated by a proposed experimental search for the electric dipole moment of the neutron (nEDM)
utilizing neutron-$^3$He capture in a dilute solution of $^3$He in superfluid $^4 $He, we derive the transport properties of 
dilute solutions in the regime where the $^3$He are classically distributed and rapid $^3$He-$^3$He scatterings keep the $^3$He in equilibrium. 
Our microscopic framework takes into account phonon-phonon, phonon-$^3$He, and $^3$He-$^3$He scatterings.
We then 
apply these calculations to measurements by Rosenbaum et al. [J.~Low~Temp.~Phys. {\bf 16}, 131 (1974)]
 and by Lamoreaux et al. [Europhys.~Lett.~{\bf 58}, 718 (2002)] of dilute solutions in the presence of a heat flow.   We find satisfactory agreement of theory with the data, serving to confirm our understanding of the microscopics of the helium in the future nEDM experiment.


\end{abstract}

\maketitle

\section{Introduction}

Dilute solutions of $^3$He in superfluid $^4$He have been an ideal testing ground for theories of quantum liquids, with past focus generally on $^3$He concentrations and temperatures for which the $^3$He forms a degenerate Fermi gas.    The proposed use of ultra-dilute solutions in the search for a neutron electric dipole moment (nEDM) at the Oak Ridge National Laboratory Spallation Neutron Source (SNS)~\cite{snsExpt,reaction}, as well as earlier experiments by Rosenbaum et al.~\cite{rosenbaum} and by Lamoreaux et al.~\cite{lamoreaux}, all require careful treatment of the transport properties of the solutions.  The characteristic temperatures in all cases considered here are of order 0.5~K, at which phonons are the dominant excitation of the $^4$He. In the low temperature degenerate $^3$He regime, phonons have small effect on the $^3$He properties.   However, 
with increasing dilution, when the $^3$He become classically distributed, the situation is reversed, and the phonons play a more and more important role.  In the proposed nEDM experiment, a crucial issue is to be able to periodically sweep out the $^3$He by imposing a temperature gradient \cite{hayden}; the underlying physics of the transport is scattering of phonons in the superfluid $^4$He against the $^3$He.  The Lamoreaux et al. experiment, which measured the effect of a heat source on the steady state distribution of $^3$He atoms in a dilute solution, was a prototype for the effects of a ``phonon wind'' on the $^3$He.  The Rosenbaum et al. experiment measured the thermal conductivity of dilute solutions as a function of temperature and concentration.  We show below that the results of these experiments can be understood in terms of a simple thermal conductivity determined using well-established scattering amplitudes.

In the  experiments of Refs.~\cite{rosenbaum} and \cite{lamoreaux}, the $^3$He number concentrations, $x_3 = n_3/(n_3+n_4)$, where $n_3$ and $n_4$ are the $^3$He and $^4$He number densities, are in the range $ 7\times 10^{-5}$ to $1.5\times 10^{-3}$ in the non-degenerate regime.  Here  momentum carried by the phonons goes primarily into the $^3$He, which then transfer it to the walls.  The physical dimensions of the experimental container are sufficiently large that the transport properties are determined locally by the microscopic scatterings of the $^3$He and the phonons.  The $^3$He-$^3$He interactions are sufficiently strong that they keep the $^3$He in thermal equilibrium at rest at the local temperature $T(\vec r\,)$, while  phonon-phonon interactions keep the phonons in drifting local equilibrium.  By contrast, in the proposed SNS experiment, phonon momentum is transferred primarily to the walls by viscous forces, with the $^3$He playing a negligible role.  Furthermore, collisions of $^3$He with the phonons are responsible for establishing equilibrium in the $^3$He cloud.   A common characteristic of these experiments is the effect on the dilute solution produced by a localized, static heat source.  In this paper we focus on calculating transport properties in the higher concentration regime in Refs.~\cite{rosenbaum} and \cite{lamoreaux}, where the fact that the $^3$He-$^3$He mean free path is short greatly simplifies the transport theory.   At lower $^3$He concentrations, $x \lesssim 10^{-6}$,  the transport must be calculated by solving the coupled Boltzmann equations for the phonons and $^3$He, taking into account viscous forces at the boundaries;  these results as well as their impact on the transport of $^3$He in the much lower concentration regime of the SNS nEDM experiment will be described elsewhere~\cite{BBPII}.  

In Sec.~II we examine the hydrodynamic constraints deriving from the steady state situation and from the properties of the superfluid.  The basic scattering mechanisms determining the transport properties of dilute solutions---$^3$He-$^3$He, $^3$He-phonon, and phonon-phonon interactions---are described in Sec.~III.  In Sec.~IV, we calculate the thermal conductivity (dominated by the phonons) in this situation; in this calculation we include inelastic recoil of the $^3$He in scattering against the phonons (detailed in Appendix A).  We find that, contrary to the earlier treatment in Ref.~\cite{BE67} used by Rosenbaum et al., the rapid relaxation of phonons along rays of constant phonon direction dominates their distribution~\cite{maris}.  We calculate the conductivity by considering the drag force on the phonons due to their scattering against the stationary $^3$He, using the method close in spirit to that introduced earlier in a calculation of the mobility of ions in superfluid $^4$He \cite{ruben}, and later used by Bowley \cite{Bowley} in his accounting of the Lamoreaux et al.~experiment.   Despite a superficial similarity of the present approach to these earlier calculations, the underlying physics is different here. 
In Sec.~V, we show how the microscopic theory satisfactorily explains the experimental findings in both the Rosenbaum et al. and the Lamoreaux et al. experiments.   

\section{Response to static disturbances}
\label{sec2}

To understand how dilute solutions respond to a localized static heat source, the physical mechanism of interest in the experiments,  we first review the hydrodynamics of the solutions.  At low temperatures the phonons are the dominant excitations of the $^4$He, and the momentum density or mass current density of the $^4$He is 
\beq
 \vec g_4 = \rho_s \vec  v_s + \rho_{ph}\vec v_{ph},
\eeq 
where $\vec v_s$ is the superfluid flow velocity, $\vec v_{ph}$ is the phonon fluid flow velocity, $\rho_s$ is the superfluid mass density, and
\beq
 \rho_{ph} = \frac{2\pi^2}{45}\frac{T^4}{s^5}\quad (\ll \rho_s)
\eeq
is the $^4$He normal fluid density.  Similarly, the $^3$He momentum density is 
\beq
  \vec g_3 =   \rho_3\vec v_3,
\eeq
where $\vec v_3$ is the $^3$He flow velocity, and  $\rho_3 = m^* n_3$, with $m^* = m_3 +\delta m \simeq 2.34 \,m_3$ the  effective mass~\cite{BE67}.   The  total mass current, $\vec g$, is $\vec g_4 + \vec g_3$.   

  When the $^4$He mass flow vanishes, $\left | v_s \right | = \left( \rho_{ph} / \rho_s \right) |v_{ph}| \ll |v_{ph}|$ at temperatures $\lesssim 0.6$~K, and the only relevant flow velocities are those of the phonons and possibly the $^3$He.   Then force balance in the dilute solutions implies that to linear order,
\beq
  \nabla P = \eta_{ph} \nabla^2 \vec v_{ph} + \eta_{3} \nabla^2 \vec v_{3},  
\label{navier} 
\eeq
where $P= P_3 +P_{ph}$ is the total pressure, with $P_3 = n_3T$ the $^3$He partial pressure (we generally work in units with $\hbar$ and Boltzmann's constant,  $k_B$,  equal to unity),  $T$ is the temperature, $P_{ph}$ is the phonon partial pressure, $\eta_{ph}$ is the first viscosity of the normal fluid and $\eta_3$ the first viscosity of the $^3$He.   (In the situations of interest, in a steady state,  $\nabla\cdot \vec v_{ph}$ and  $\nabla\cdot \vec v_{3}$ both vanish.)  For a container large compared with microscopic viscous mean free paths, and for $^3$He concentrations in the range of those in Refs.~\cite{rosenbaum}  and \cite{lamoreaux}, both viscosity terms are insignificant compared with the drag forces between the $^3$He and the phonons, as we discuss in Sec.~\ref{sec:results}, and thus can be neglected. (At the much lower $^3$He concentrations of the proposed SNS experiment, however, the phonon viscosity does play a significant role~\cite{BBPII}.)  The total pressure is effectively constant throughout the system, $\nabla P = 0$.    

In addition, as one sees from the linearized superfluid acceleration equation~\cite{khalat},
\beq
m_4\frac{\partial \vec v_s}{\partial t} +\nabla \mu_4 = 0,
\eeq
the $^4$He chemical potential, $\mu_4$, is  constant in a steady state.  The 
Gibbs-Duhem relation for the solutions, $\nabla P = n_4 \nabla \mu_4 + n_3 \nabla \mu_3 + S \nabla T$,  with  $\mu_3$ the $^3$He chemical potential, $T$ the temperature, and $S$ the total entropy density,  together with the constancy of the pressure and the relation, $\nabla P_3 =  n_3 \nabla \mu_3 + S_3 \nabla T$ (which neglects the effects of $^3$He-phonon interactions on the thermodynamics,  e.g., small terms in the total pressure of order $P_{ph}n_3/n_4\ll P_{ph}$),
then implies that
\beq
\nabla P_3 + S_{ph} \nabla T = 0, 
\label{p3t}
\eeq
where 
$S_{ph} = 4P_{ph}/T$ is the phonon entropy density, and $dP_{ph} = S_{ph}dT$.  Equation (\ref{p3t}) and the foregoing then give the simple relation between the temperature and $^3$He density gradients in the system:
\beq
\nabla T = -\frac {T}{S_{ph} + n_3} \nabla n_3; \label{eqn:Tn3}
\eeq
in a steady state a gradient of the $^3$He density is always accompanied by a gradient of the temperature.  (Note that the $n_3$ in the denominator arises when one consistently includes a non-zero temperature gradient at every step of the calculation, unlike in earlier studies \cite{khalat2,wilks} where the $^3$He pressure was tacitly assumed to obey $dP_3 = Tdn$.)  
  A heat flux, $\vec Q$, in the system is related to the temperature gradient by 
$\vec Q  = -K \nabla T, $   where $K$ is the thermal conductivity of the solution.    Thus Eq.~(\ref{eqn:Tn3}) relates the $^3$He density gradient to the heat flux by
\beq
  \nabla n_3 = \frac{S_{ph}+n_3}{TK}\vec Q.
  \label{n3q}
\eeq
The  $^3$He density on the right can be significant.  Since 
\beq
 TS_{ph}  = s^2 \rho_{ph}
\label{eqn:Sph}
\eeq
one has
\beq
 \frac{n_3}{S_{ph}} = 300 \frac{x_3}{T^3},
\eeq
with $T$ measured in K; at $T$ = 0.45 K and $x_3 = 3\times 10^{-4}$ the ratio is unity.

We note that, in general, the response to a heat current in a steady state is a temperature gradient with the constant of proportionality being the thermal conductivity; the particle current in general depends on both gradients of concentration and temperature~\cite{LL}.  In the present case, because of the relation between $\nabla n_3$ and $\nabla T$, Eq.~(\ref{eqn:Tn3}), the $^3$He diffusion coefficient is proportional to the thermal conductivity and thus the results of Ref.~\cite{lamoreaux} can also be described in terms of a diffusion coefficient.  However, when viscous effects become important (for $x_3 < 10^{-6}$), Eq.~(\ref{eqn:Tn3}) is no longer valid and the thermal conductivity and diffusion constant are not related in a simple way~\cite{BBPII}.  In the following we will generally talk about the response to a heat current in terms of the thermal conductivity.

\section{Microscopic scattering processes  \label{sec:microscopic}}
 
The microscopic processes that determine the transport properties of dilute solutions are $^3$He-$^3$He \cite{BE170}, phonon-phonon \cite{maris}, and phonon-$^3$He \cite{BE67} scatterings; the amplitudes of these processes have been well studied in earlier helium research.  At low energies, the total cross section for a $^3$He atom scattering from a second $^3$He atom of opposite spin can be written as 
\beq
\sigma_{33} = \frac{9\pi^3}{k_D^2}v_{33,0}^2\left( \frac{m^*}{m_4} \right )^2 \left (\frac{m_4 s}{k_D} \right )^4 = 10.3 \rm\,\AA^2,
\eeq
where $k_D$ is the Debye wave number, defined by $n_4 = k_D^3/6 \pi^2$, and $v_{33,0} = -0.064$ measures the strength of the $^3$He-$^3$He interaction at zero momentum transfer~\cite{BE170}.   Phonon-phonon scattering rates at forward angles~\cite{maris} are rapid compared with phonon-$^3$He rates.  While such scatterings do not directly affect the heat current, they do play the important role of keeping the phonon momentum distribution, $n_{\vec q}$, in local thermodynamic equilibrium with a drift velocity, $\vec v_{ph}$,
\beq
  n_{\vec q} = \frac{1}{e^{(sq-\vec q\cdot \vec v_{ph})/T}-1}.
\eeq  
This is the key difference between the current treatment and that in Refs.~\cite{rosenbaum} and \cite{BE67}, where it is assumed that the phonon-$^3$He scattering dominates the phonon relaxation.

   The amplitude for scattering of a phonon of wavevector $\vec q$ against a  $^3$He of wavevector $\vec p$ to final states $\vec q\,'$ and $\vec p\,'$  is determined theoretically in terms of the measured excess volume, $\alpha$, of a $^3$He atom compared with that of a $^4$He atom and the $^3$He effective mass, $m^*$.  The scattering amplitude is approximately \cite{BE67}
   \beq
  \langle p'q'|{\cal T}|pq\rangle = \frac{s(qq')^{1/2}}{2 n_4} (A+B\cos \theta),
  \label{scatamp}
\eeq  
where $\theta$ is the angle between $\vec q\,'$ and $\vec q$,
$A= n_4 d\alpha/dn_4$ and $B= (1+\alpha +\delta m/m_4)(m_4/m^*)(1+\alpha - m_3/m_4)$. 
   
At very low temperatures, $T \ll 1$~K, the scattering can taken to be elastic.  The momentum dependent scattering rate of phonons from the $^3$He is then~\cite{BBPII}
\beq
\gamma_3\left( q \right) = \frac{sq^4 x_3}{4 \pi n_4}J,
\label{gamma}
\eeq
where 
\beq
J = A^2 + B^2/3 -2AB/3;
\eeq 
the $q^4$ in this rate is characteristic of Rayleigh scattering.  
More generally, as Bowley emphasized \cite{Bowley},  recoil of the
$^3$He in scattering produces a significant correction to the effective
phonon-$^3$He scattering rate~\cite{bowleyError}.  As we show in Appendix A, $^3$He recoil corrections to the result (\ref{gamma}) are of relative order $T/1.36$~K.

  We now estimate the importance of 3-3 versus phonon scattering in bringing the $^3$He into equilibrium.  
The mean free path of a $^3$He scattering on unpolarized $^3$He is
\beq
\ell_{33}=\frac{2}{n_3 \sigma_{\rm 33}}  =\frac{8.66\times 10^{-8}}{x_3} {\rm cm};
\label{l33} 
\eeq
the factor $n_3/2$ is the density of opposite spin $^3$He.   Similarly, the mean free path for scattering of $^3$He of momentum $p$ on phonons is given approximately by $p^3/m\Gamma$, where
\beq
\Gamma &=& \frac{1}{n_3} \int \frac{d^3 q}{\left ( 2 \pi \right )^3} q^2 \gamma_3 \left( q \right) n_q^0 \left ( 1 + n_q^0 \right ) \nonumber\\
&=& 270 \pi J  \left(\frac{S_{ph}}{n_4}\right)^2    \frac{T^3}{s^2} \sim T^9,
\label{Gamma}
\eeq
in the limit $p \gg q$.  Here $n_q^0 = \left ( e^{sq/T} - 1 \right )^{-1}$ is the equilibrium phonon distribution function. 
Replacing $p^2$ by $3m^*T$, an appropriate thermal average, we find
\beq
  \ell_{3ph} &=&   \frac{\sqrt3}{2J}  \left(\frac{S_{ph}}{n_4}\right)^2   \frac{m^{*1/2} s^2}{T^{3/2}} 
    \nonumber \\
      &=& 0.077 \left(\frac{0.45\,K}{T}\right)^{15/2} \, {\rm  cm}.
 \label{l3ph}     
\eeq  
Comparing the mean free paths ({\ref{l33}) and (\ref{l3ph}), we find
\beq
\frac{\ell_{3ph}}{\ell_{33}} =   0.89 \times 10^6 x_3   \left(\frac{0.45\,K}{T}\right)^{15/2}.
\label{ellratio}
\eeq 
For $T=0.45$~K and $x_3=10^{-6}$, $\ell_{3ph} \approx \ell_{33}$, while for 
$T=0.65$~K and $x_3=3\times 10^{-4}$,  $\ell_{3ph}/\ell_{33}
\approx  16.9.$
Thus for $T\sim 0.5$~K, and $x_3\sim 10^{-5}$ a good first approximation is to assume that scattering of $^3$He by $^3$He atoms is fast compared with scattering of $^3$He by phonons.  In fact, as will be shown in detail in \cite{BBPII}, we may take the momentum distribution of the $^3$He  to be simply that
of a classical gas in equilibrium and at rest,
\beq
f_p^0 = e^{-(p^2/2m^* -\mu_3)/T}.
\eeq

\section{Thermal conductivity}
\label{sec:results}
The thermal conductivity of the dilute solutions can in general be calculated by solving the 
coupled phonon and $^3$He Boltzmann equations to determine the steady state momentum distributions of the phonons and the $^3$He, and, from these distributions, the heat and particle currents.  However, here, where the $^3$He are approximately stationary and in equilibrium, the phonon thermal conductivity can be found simply, by calculating the rate at which the phonons lose momentum by scattering against $^3$He \cite{ruben,Bowley}. 

In a steady state the net force density on the phonons drifting at velocity $\vec v_{ph}$ 
 is balanced by the phonon pressure gradient, $\nabla P_{ph} = S_{ph}\nabla T$.   Microscopically then,
\beq
\nabla P_{ph} = &&-\int \frac{d^3q}{(2\pi)^3}  \int \frac{d^3q'}{(2\pi)^3}\int 2 \frac{d^3p}{(2\pi)^3} \nonumber\\
&&\hspace{-18pt}\times \vec q\,|\langle {\cal T}\rangle|^2 2\pi\delta(sq+p^2/2m^* -sq' -p'^2/2m^*) \nonumber\\
 && \hspace{-16pt} \times \left[f^0_{p}n_{\vec q}\,(1+n_{\vec q\,'})-f^0_{p\,'}n_{\vec q\,'}(1+n_{\vec q})\right], \nonumber\\
 \label{phBE}
\eeq
where $ \vec p\,'-\vec p = \vec q - \vec q\,' \equiv \vec k$, $\langle {\cal T}\rangle = \langle p'q'|{\cal T}|pq\rangle$, $n_{\vec q}$ is the phonon distribution, and the factor of 2 is from the $^3$He spin sum.  In terms of the $^3$He structure function,
\beq
  S_3(k,\omega) =&& 2\int \frac{d^3p}{(2\pi)^3} \delta(\omega+p^2/2m^* -p'^2/2m^*)\, f^0_{p} ,\nonumber\\
     =&& n_3\left(\frac{m^*}{2\pi k^2T}\right)^{1/2} e^{-m^*(\omega - k^2/2m^*)^2/2k^2T}, \nonumber\\
 \label{S3withrecoil}    
\eeq  
which obeys $ S_3(k,-\omega)= e^{-\omega/T}S_3(k,\omega)$, we can rewrite Eq.~(\ref{phBE}) as,
\beq
\nabla P_{ph} = &&-\int \frac{d^3q}{(2\pi)^3}  \int \frac{d^3q'}{(2\pi)^3}  \int_{-\infty}^{\infty} d\omega \nonumber\\
&&\hspace{-18pt}\times  \vec q\,|\langle {\cal T}\rangle|^2 2\pi\delta(\omega - sq+sq')   S_3(k,\omega) \nonumber\\
 && \hspace{-16pt} \times\left[ n_{\vec q}\,(1+n_{\vec q\,'})-n_{\vec q\,'}(1+n_{\vec q})e^{-\omega/T}\right]. \nonumber\\
 \label{phBE2}
\eeq
Since $n_{\vec q\,'}(1+n_{\vec q}) = n_{\vec q}\,(1+n_{\vec q\,'})e^{(\omega - \vec k\cdot \vec v_{ph})/T}$, the final line in Eq.~(\ref{phBE2}) to first order in $v_{ph}$ is 
$n_q^0(1+n_{q'}^0)(\vec k \cdot \vec v_{ph})/T$.  Finally, symmetrizing the integrand in $q$ and $q'$, and carrying out the angular averages, we have, in agreement with Bowley,
\beq
\nabla P_{ph} = &&-\frac{\vec v_{ph}}{6T}\int \frac{d^3q}{(2\pi)^3}  \int \frac{d^3q'}{(2\pi)^3} n_q^0 (1+n_{q'}^0) k^2  |\langle {\cal T}\rangle|^2
\nonumber\\
&&\hspace{-6pt}\times\int_{-\infty}^{\infty} d\omega  \,2\pi\delta(\omega - sq+sq')   S_3(k,\omega). 
 \label{phBE3}
\eeq

   Neglecting $^3$He recoil is equivalent to setting $S_3(k,\omega) = n_3\delta(\omega)$.  In this case,
\beq   
   \nabla P_{ph} =  &&-\frac{\vec v_{ph}}{6T}\int \frac{d^3q}{(2\pi)^3}  \int \frac{d^3q'}{(2\pi)^3} n_q^0 (1+n_{q'}^0) k^2  |\langle {\cal T}\rangle|^2
\nonumber\\
&&\hspace{36pt}\times  2\pi\delta(sq-sq') , \nonumber\\
=&&-\frac{n_3\Gamma}{3T}\vec v_{ph} =-\frac{8\pi^5}{45}\frac{x_3J}{n_4}\left(\frac{T}{s}\right)^8 \vec v_{ph} .
 \label{phBE34}
\eeq
The phonon heat current density, 
\beq
 \vec Q_{ph} = s^2\int\frac{d^3q}{(2\pi)^3} \vec q \, n_{\vec q} = TS_{ph} \vec v_{ph},
 \label{qph}
 \eeq
in steady state determines the phonon thermal conductivity, $\vec Q_{ph} = -K_{ph}\nabla T$, and
since $\nabla P_{ph} = S_{ph}\nabla T$, we find the simple result~\cite{xto0}
\beq
K_{ph} = S_{ph}s\ell_{ph3}=  \frac{3 T^2 S_{ph}^2}{n_3 \Gamma}
  =\frac{n_4 s^2}{90 \pi x_3 JT},
\label{kph}
\eeq
where $\ell_{ph3}=  3 T^2 S_{ph}/sn_3 \Gamma$ defines an effective mean free path for phonons scattering on $^3$He.

 In Appendix A we include $^3$He recoil to leading order in $T/m^*s^2$,  which we find increases  the thermal conductivity by $\sim
25-35$\% in the range $T=0.45-0.65$~K.
Figure~\ref{fig:K} shows the phonon thermal conductivity 
computed from Eq.~(\ref{Kphrecoil}), as well as the approximate thermal conductivity,
 Eq.~(\ref{kph}) together with the measurements of Ref~\cite{rosenbaum}.  We use here the parameters $A= -1.2\pm0.2$ \cite{BM67,WRR69}, and $B= 0.70 \pm 0.035$ \cite{BM67,WRR69,A69}, which lead to  $J= 2.2 \pm 0.6$. The largest uncertainty is in $A$, owing to a systematic difference between the measurements \cite{BM67,WRR69} of the pressure dependence of the density of dilute solutions.

\begin{figure}
\includegraphics[width=8.5cm]{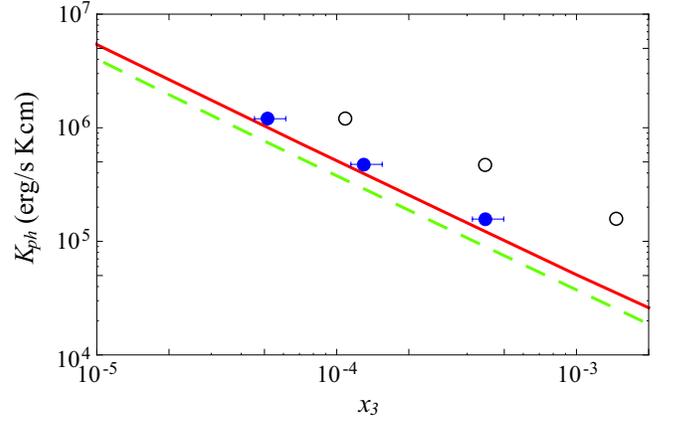}
\caption{(Color online) Thermal conductivity of a dilute solution of $^3$He in superfluid $^4$He at $T$ = 0.45 K.  
The solid line shows the phonon thermal conductivity, in the limit in which the $^3$He are at rest in equilibrium, calculated with $^3$He recoil (see text).  The dashed line shows
the phonon thermal conductivity, Eq.~(\ref{kph}), without recoil corrections.  The open circles are three representative measurements 
at $T \sim 0.43$~K at the average $x_3$
reported in Ref~\cite{rosenbaum}; the solid circles with the error bars are the same thermal conductivities, but shown at the corrected $x_3$, as discussed in Sec.~\ref{sec:comparison}.
 \label{fig:K}}
\end{figure}

 \begin{figure}
\includegraphics[width=8.5cm]{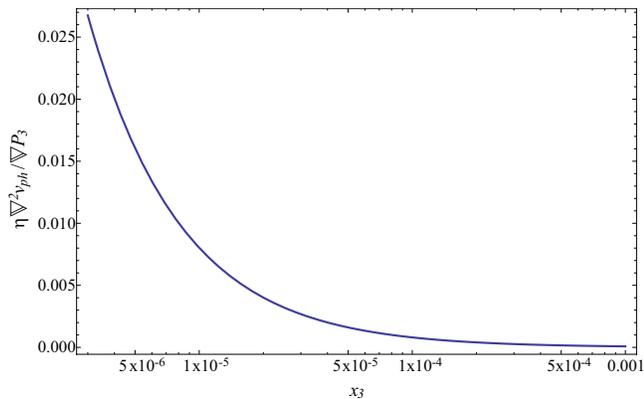}
\caption{Ratio of the contribution of the viscous term (see Eq.~(\ref{navier})) to the drag on the phonons due to scattering against the $^3$He at a temperature of 0.45 K.  The magnitude of the drag is given by the $^3$He pressure gradient (see Eq.~(\ref{phBE34})) as a function of $^3$He concentration $x_3$ (the concentrations measured in~\cite{lamoreaux} are $x_3 = 7\times 10^{-5}$ and $3\times 10^{-4}$).  The viscous term is calculated for laminar flow in a circular pipe of radius $R$ = 2.5 cm, using the $^4$He viscosity measured by Greywall~\cite{greywall}. }
\label{fig:visc}
\end{figure}

    We return now to justify neglecting viscous stresses in the pressure equation (\ref{navier}).  The drag force per unit volume of the $^3$He on the phonons is $\sim \rho_{ph}s/\ell_{ph3}$.   We
    assume  laminar flow in a circular pipe of radius $R$, for which  $\nabla^2 \vec v_{ph} = -8\langle \vec v_{ph}\rangle /R^2$ is a constant throughout \cite{visc}, thus overestimating the viscous force (density) on the phonons by neglecting the drag against the $^3$He.  The phonon viscosity is $\sim \rho_{ph}s\ell_{visc}/5$, where $\ell_{visc}$ is the phonon mean free path for viscosity, and thus the ratio of the viscous force to drag force is $\sim \ell_{visc}\ell_{ph3}/R^2 \ll 1$; see Fig. \ref{fig:visc}.  In addition, the ratio of the $^3$He viscosity,  $\sim m^*n_3 \bar v\ell_{33}/5$ (where $\bar v$ is a mean $^3$He thermal velocity), to the phonon viscosity is of order $(n_3/S_{ph})(\ell_{33}/\ell_{visc}) \ll 1 $ here.
    
\section{Diffusion of $^3\mbox{\boldmath He}$ against phonons}
\label{sec:diffusion}  

   We now calculate the rate of diffusion of $^3$He atoms in a bath of phonons at a fixed temperature $T$.   In the derivation of heat transport above, we assumed that the $^3$He cloud was at rest and that the phonons were drifting with  velocity $\vec v_{ph}$.   In a frame in which the phonons are at rest and the $^3$He drift with velocity $\vec v_3$,
 Eq.~(\ref{phBE34}) becomes
 \beq   
   \nabla P_{3} = -\frac{n_3\Gamma}{3T}\vec v_{3},
 \label{phBE35}
\eeq 
where we use $\nabla P_{ph} = -\nabla P_3$.  With the temperature variation in $P_3$ neglected, Eq.~(\ref{phBE35}) becomes a familiar diffusion equation,
\beq  
    D_3\nabla n_3 = - n_3 \vec v_{3},
 \label{d3a}
\eeq
where 
\beq
  D_3 = \frac{3T^2}{\Gamma} \sim \frac{1}{T^7}.
  \label{D}
\eeq

More generally, however, phonons drive $^3$He pressure gradients, rather than simply density gradients.  In an arbitrary frame, the response of the $^3$He current to a phonon wind becomes
 \beq
  \vec j_3 = n_3 \vec v_3 = n_3 \vec v_{ph} -D_3\nabla n_3 -  D_T\nabla T,
  \label{j3}
 \eeq
where, in our particular case, $D_T = n_3D_3/T$ is an effective ``thermoelectric" transport coefficient.

\section{Application to experiment}
\label{sec:comparison}

To further illustrate the physics, we now determine the temperature and concentration distributions for two geometries (Refs.~\cite{rosenbaum} and \cite{lamoreaux}) of interest.  Conservation of energy implies
\beq
\nabla \cdot \vec Q = -\nabla \cdot (K_{ph}\nabla T) = -C\nabla \left(\frac{\nabla T}{P_3}\right) =0,
\label{eqn:divQ}
\eeq
where, from Eq.~(\ref{kph}), $C= n_4^2 s^2/90\pi J$.  For simplicity, recoil effects are neglected in writing the equations here; however, they are taken into account in all numerical calculations reported below.  We eliminate $P_3$ using the solution
\beq
P_3 + \frac14S_{ph}T = P,
\label{eqn:P3Soln}
\eeq
of Eq.~(\ref{p3t}), where the constant $P$ is the total pressure of the excitations.   Equation~(\ref{eqn:divQ}) then reduces to a partial differential equation in $T$ alone (even when including the recoil corrections), which we solve using the finite element code FlexPDE~\cite{flexpde}.   Given $T(\vec r)$, we determine $n_3(\vec r)$ from Eq.~(\ref{eqn:P3Soln}), and determine the constant $P$ by fixing the total number of $^3$He atoms in the system.

We begin by applying this theoretical description to the experiment of Rosenbaum et al.~\cite{rosenbaum}, where the thermal conductivity of mixtures in the concentration range $1.1\times 10^{-4} \le x_3 \le 1.3\times 10^{-2}$ were measured at temperatures $0.084 \le T \le 0.65$~K.  They determined the thermal conductivity by measuring the temperature difference over a 5 cm length of pipe, 0.26 cm in diameter, in the presence of a localized heat source.  The pipe was connected to small reservoirs at either end containing the thermometers, the coupling to the dilution refrigerator and the heater.  

In the analysis in Ref.~\cite{rosenbaum}, the effect of the variable $^3$He concentration in the pipe, and its attendant effect on the thermal conductivity, was not taken into account.  We therefore determined the temperature and concentration distributions in their system by adjusting the heater power and minimum (dilution refrigerator) temperature to match the reported average temperature and the temperature difference implied by the reported thermal conductivity.  We note that the $^3$He thermal conductivity is negligibly small owing to $\ell_{33}\ll \ell_{3ph}$, Eq.~(\ref{ellratio}).

Because the phonons push the $^3$He into the cold reservoir, the average concentration in the pipe is substantially lower than the concentration including the reservoirs.  We show the results of our calculations in Fig.~\ref{fig:K} for representative measurements~\cite{tempRange} of Ref.~\cite{rosenbaum}.  The open circles are the thermal conductivities as reported, plotted at the average overall concentration; the solid circles are plotted at the average concentrations calculated for the pipe.  Because the dimensions of the reservoirs are not given in Ref.~\cite{rosenbaum}, there is some uncertainty in the calculated result.  The error bars shown in Fig.~\ref{fig:K} represent changes of about $\pm 25$\% from the reservoir volumes estimated from the schematic in their Fig.~1.  Given the uncertainties related to the geometry and the details of how the thermal conductivity was extracted, there is good agreement between this calculation and the measurement.  As previously noted in Sec.~\ref{sec:microscopic}, the theoretical analysis in Ref.~\cite{rosenbaum} incorrectly assumed that the primary phonon relaxation was due to $^3$He-phonon scattering, rather than phonon-phonon scattering along rays.

We now analyze the experiment of Lamoreaux et al.~\cite{lamoreaux}, which measures the $^3$He density response to a localized heat source by a novel technique. In the experiment, a dilute solution with concentration in the range $7\times 10^{-5}$ to $3\times 10^{-4}$
is contained in a cylindrical cell roughly 5 cm in diameter and 5 cm long, cooled at one end by a dilution refrigerator to a temperature $T$ in the range $0.45-0.95$~K~\cite{hayden2}.   A temperature gradient is created by a resistive heater generating  7--15 mW, located roughly midway between the ends and near the cylinder wall (see Fig.~\ref{fig:x3Contour} below).  The resulting spatial distribution of the $^3$He density is probed by a well-collimated neutron beam of diameter $\sim$ 0.25 cm going through the cell.  A fraction of the neutrons are captured via the reaction ${\rm n} + ^3{\rm He} \rightarrow {\rm p} + {\rm t} + 764 \hbox{ keV}$.
The XUV scintillation light from protons and tritons in the liquid $^4$He is converted to visible light and detected by a photomultiplier tube which views the solution through a window at the other end of the cell.  The yield of scintillation light, measured as a function of cell temperature, initial $^3$He concentration, and heater power, is used to determine the thermally induced change in the $^3$He distribution.  The heat flow again produces non-uniform temperature and $^3$He concentration distributions.  Because the concentration and temperature gradients are directly proportional, Eq.~(\ref{eqn:Tn3}), the concentration gradients implied by the measured scintillation yields here simply encode the same type of temperature difference information measured by Rosenbaum et al.  We note that the proportionality constant involves $S_{ph}+n_3$, rather than just $S_{ph}$.
 
\begin{figure}
\includegraphics[width=8cm]{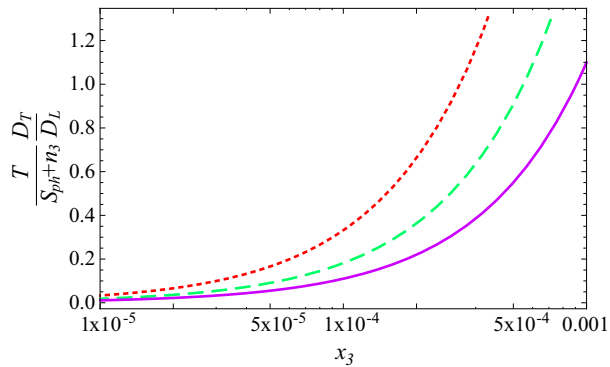}
\caption{(Color online) Contribution of the thermoelectric transport term relative to the effective diffusion constant, $D_L$ extracted in Ref.~{lamoreaux}.  The temperatures are 0.45~K (dotted), 0.55~K (dashed) and 0.65~K (solid).
\label{fig:DT}}
\end{figure}

\begin{figure}
\includegraphics[width=8cm]{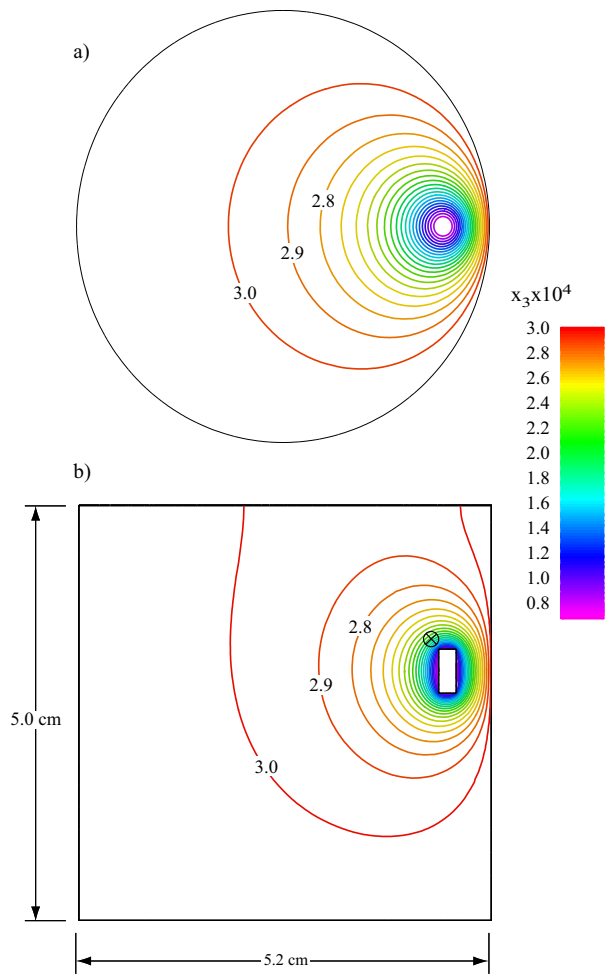}
\caption{(Color online) An example of the $^3$He distribution for a cell 5.2~cm in diameter and 5.0~cm high, with refrigerator temperature 0.45~K, average $x_3 = 3\times10^{-4}$, and heater power 15 mW.  The sections are: a) perpendicular to the cylinder axis and containing the neutron beam axis; and b) perpendicular to the neutron beam (the cross in the figure) and through the center of the heat source.  We assume in this simulation that the cylindrical wall of the cell is at the same temperature as that of the base where the refrigerator is attached; the top surface is insulated.  The $^3$He concentration contours, of constant spacing, are marked in units of $10^{-4}$; the minimum is $x_3 = 0.69\times10^{-4}$ at the surface of the heater and the maximum concentration is $3.05\times10^{-4}$, on the lower cell boundary. 
{\label{fig:x3Contour}}}
\end{figure}

\begin{figure}
\includegraphics[width=8cm]{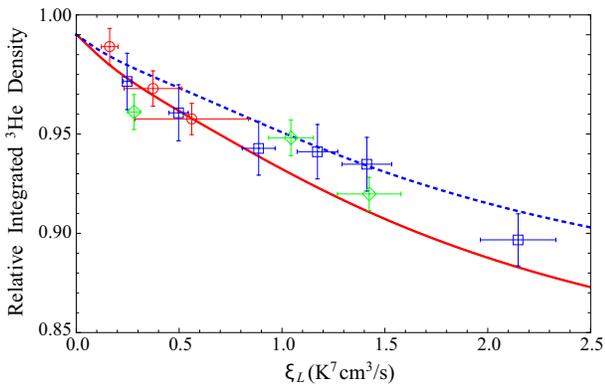}
\caption{(Color online) Representative results of the finite element calculation, including $^3$He recoil, of the relative integrated $^3$He (column) density along the neutron beam as a function of the quantity $\xi_L = T^7 \left ( {\cal P}/S_{ph}T \right )$ where ${\cal P}$ is the heater power, along with data from Ref.~\cite{lamoreaux} (circles, 0.45~K; diamonds, 0.55~K and squares, 0.65~K).  The curves are for a refrigerator (and barrel) temperature of $0.45$~K: {\em solid}, $x_3 = 3\times 10^{-4}$;  {\em dotted}, $x_3 = 7\times 10^{-5}$; coincidentally, the results for $T=0.65$~K and $x_3 = 3\times 10^{-4}$ are essentially the same as the dotted curve.  
The curves are plotted for the average temperature (greater than the boundary temperature) along the neutron beam path. For example, at the actual value $T^6 {\cal P}/S_{ph}  = 2.5$, the average $\xi_L$ is shifted upward by $\Delta \xi_L \approx 0.05\,K^7$ cm$^3$/s for $T$ = 0.45 K, and by $\Delta \xi_L \approx 0.5\,K^7$ cm$^3$/s for $T = 0.65$~K.
{\label{fig:dataComparison}}}
\end{figure}

The response of the $^3$He density to the heat source is given here by Eq.~(\ref{n3q}).
Combining this result with $\vec Q = T S_{ph} \vec v_{ph}$,  one can write
\beq
\nabla n_3 = \frac{S_{ph}(S_{ph}+n_3)}{K}\vec v_{ph}.
\label{eqn:n3K}
\eeq 
Reference \cite{lamoreaux}, which did not take into account the temperature gradient induced by the heat flow, interpreted the result for the $^3$He density gradient in terms of a simple diffusion constant, $D_L$, in the form,
\beq
\nabla n_3 = \frac{n_3}{D_L} \vec v_{ph}.
\eeq   
From Eq.~(\ref{eqn:n3K}) we see that 
\beq
D_L &=& \frac{n_3}{S_{ph} + n_3}\frac{K}{S_{ph}}.
\eeq
Equivalently,
\beq
  D_L= D_3 - \frac{T}{S_{ph}+n_3} D_T = \frac{S_{ph}}{S_{ph}+n_3} D_3.
\label{eqn:DLDph3DT}
\eeq
The $D_T$ term in this equation makes a significant contribution at higher concentrations (see Fig.~\ref{fig:DT});
as noted, $n_3 \sim S_{ph}$ for a large range of concentrations in the experiment of Ref.~\cite{lamoreaux}.  This effective $D_L$ is not a simple diffusion constant,  owing to the presence of the temperature gradient. 
We note that while the temperature dependence of the result for $D_3$ without $^3$He recoil falls with temperature as $1/T^{7}$, the temperature dependence of the effective $D_L$ differs, owing both to recoil effects and the $1/T^3$ dependence of $n_3/S_{ph}$.  The relative similarity of $D_L$ and $D_3$ also depends on the $^3$He thermal conductivity being negligible.

An example of the $^3$He distribution resulting from the finite element calculation for the Lamoreaux et al. experiment is shown in Fig.~\ref{fig:x3Contour}; results of the relative integrated (column) $^3$He densities along the neutron beam are shown in Fig.~\ref{fig:dataComparison} together with the data in Fig. 4 of \cite{lamoreaux}.  To illustrate the effect of other variables in the problem, we also show in Fig.~\ref{fig:dataComparison} the results for $T$ = 0.65~K and for the two concentrations used in the experiment.  In this calculation we take the refrigerator end and the barrel of the cell to be fixed at $T=0.45$ and $0.65$~K as indicated; the opposite end of the cell is insulated.  We note that the conductivity of the aluminum barrel is more than an order of magnitude larger than that of the phonons at these temperatures~\cite{AlCond}.  The relative column density calculated with an insulator in place of the aluminum barrel (not shown) falls well below the data.  In this simulation we neglect any possibility of convective flow in the $^3$He. The effect of the non-zero size of the neutron beam is to reduce the calculated ratios in Fig.~\ref{fig:dataComparison} by $\sim 0.003$ at the largest values of $\xi_L$, well within the reported uncertainties.  The agreement of the present theory with the experiment provides further confirmation of the microscopic transport theory in the concentration range of the Lamoreaux et al. experiment.

\section{Summary\label{sec:summary}}

 We have laid out the basic transport theory of dilute solutions of $^3$He in superfluid $^4$He in the regime where scattering among the $^3$He is the primary mechanism keeping the $^3$He in thermal equilibrium, and phonons are the significant excitations of the $^4$He. 
We find that in the range of $^3$He concentrations in the Rosenbaum et al.~\cite{rosenbaum} and Lamoreaux et al.~\cite{lamoreaux} experiments, $7\times 10^{-5} \le x_3 \le 1.5\times 10^{-3}$,  heat transport is dominated by phonons.  The physical response of the system is not simple diffusion, described only by a $^3$He diffusion constant, $D_{3}$, since the temperature gradients are important.  The experiments can be characterized simply by a phonon thermal conductivity.  This thermal conductivity, calculated in a microscopic framework, satisfactorily reproduces the measurements, indicating that the well-tested theory of $^3$He-phonon scattering in dilute solutions of $^3$He in superfluid $^4$He is consistent with these experiments as well.

\section*{Acknowledgments\label{sec:acknowledgements}}

This research was supported in part by NSF Grants PHY-0701611, PHY-0855569, PHY-0969790 and PHY-1205671.  We thank the authors of \cite{lamoreaux}, especially Robert Golub, Michael Hayden and Jen-Chieh Peng, for helpful discussions about the experiment.  GB is grateful to the Aspen Center for Physics, 
supported in part by NSF Grant PHY-1066292, and the Niels Bohr International Academy where parts of this research were carried out.  DB thanks Caltech, under the Moore Scholars program, where parts of this research were carried out.

\appendix
\section{Recoil corrections}

In this Appendix  we calculate contributions to the thermal conductivity
due to the finite $^3$He mass.  We start from Eq.\ (\ref{phBE3})
with the structure factor (\ref{S3withrecoil}) of the $^3$He,
\begin{widetext}
\beq
\nabla P_{ph} = &&-\vec v_{ph}\frac{2\pi n_3}{6T}\int
\frac{d^3q}{(2\pi)^3}  \int \frac{d^3q'}{(2\pi)^3} n_q^0 (1+n_{q'}^0)
k^2  |\langle {\cal T}\rangle|^2
  \left(\frac{m^*}{2\pi k^2T}\right)^{1/2} e^{-m^*(sq-sq' -
k^2/2m^*)^2/2k^2T}. \nonumber\\
  = &&-\vec v_{ph}\frac{2\pi n_3}{6T}\int \frac{d^3q}{(2\pi)^3}  \int
\frac{d^3q'}{(2\pi)^3}  \frac{k^2 |\langle {\cal
T}\rangle|^2}{4\sinh(sq/2T)\sinh(sq'/2T)}
   \left[\left(\frac{m^*}{2\pi k^2T}\right)^{1/2}
e^{-m^*(sq-sq')^2/2k^2T}\right] e^{ - k^2/8m^*T}.\nonumber \\
\label{phBE3Appendix}
\eeq
\end{widetext}
The integral in Eq.\ (\ref{phBE3Appendix}) is proportional to the
inverse of the thermal conductivity.   In the limit $m^*\rightarrow
\infty$ the expression in square brackets reduces to $\delta(sq-sq')$.  For finite $m^*$ there are two effects.  First, the structure factor,
and therefore also  the scattering rate, is reduced in magnitude
because of the nonzero momentum transfer, $\vec k=\vec q-{\vec q}'$, as
is shown by the final Gaussian factor. Second, as the first Gaussian
factor indicates, there is an energy transfer $sq-sq'$ which is of order
$(Tk^2/m^*)^{1/2} \sim (T/m^*s^2)^{1/2}T$. 

The contributions to the scattering amplitude for nonzero energy
transfer and for nonzero velocity of the $^3$He atoms have not been
investigated in detail, although the basic processes were discussed in full in
Ref.\ \cite{BE67}. Here we use Eq.~(\ref{scatamp}) for the scattering amplitude.  With prefactors omitted, the quantity to be calculated is thus
\begin{widetext}
\beq
\int_0^\infty dq  \int_0^\infty dq' \int_{-1}^1 d\cos \theta \frac{k
(qq')^3(A+B\cos \theta)^2}{4\sinh(sq/2T)\sinh(sq'/2T)}
e^{{-m^*s^2(q-q')^2/2k^2T}} e^{- k^2/8m^*T}.
\label{phBE3Appendix2}
\eeq
\end{widetext}
We have evaluated the integrals numerically and find, as stated in Sec.\
\ref{sec:results},  that inclusion of recoil effects increases the thermal conductivity by $\sim
25-35$\% in the temperature range $0.45-0.65$ K.

The leading corrections to the result for $m^*\rightarrow \infty$ are of
relative order $T/m^*s^2$ relative to the result in the low temperature limit;
we now calculate them analytically.  To first order in $T/m^*s^2$ the
effects of the nonzero momentum transfer and the nonzero energy transfer
are additive, and we calculate each in turn.  The more
important term is due to the momentum transfer.  When this term alone
is included one finds
\beq
{\lim_{T\rightarrow 0}(TK)}/{TK}
\simeq 1-\frac{1}{8m^*T}\frac{\langle k^4\rangle}{\langle k^2\rangle},
\eeq
where
\beq
\langle\ldots\rangle =\int_0^\infty dq \int_{-1}^1 d\cos \theta \frac{
q^6(A+B\cos \theta)^2}{4\sinh^2(sq/2T)}(\ldots) \,\,.
\eeq
In these integrals we may replace $k^2$ by its value $2q^2(1-\cos
\theta)$ for zero energy transfer.  The integrals over $q$ and $\theta$
decouple and one finds
\beq
TK \simeq \left(
1+\frac{T}{4m^*s^2}\frac{\tilde J}{J} 
\frac{I_{10}}{I_8}\right)
\lim_{T\rightarrow 0}(TK),
\eeq
where
\beq
{\tilde J}=\int_{-1}^1\frac{d\cos \theta}{2} (A+B\cos
\theta)^2(1-\cos\theta)^2\nonumber \\= \frac43 A^2 - \frac43 A B +
\frac8{15} B^2
\eeq
and
\beq
I_n=\int_0^\infty dq \frac{x^n}{4\sinh^2(x/2)} =n! \,\zeta(n),
\eeq
where $\zeta(n)$ is the Riemann zeta function of order $n$.
Therefore the thermal conductivity is given by
\beq
K \simeq \left(1+\frac{25\pi^2}{11}   \frac{\tilde J}{J}
\frac{T}{m^*s^2}\right)
\frac{1}{T}
\lim_{T\rightarrow 0}(TK).
\eeq

We now calculate the leading correction to the thermal conductivity due
to the energy transfer, which is found by neglecting the term
$k^2/8m^*T$ in the exponent in Eq.\ (\ref{phBE3Appendix2}).  The
Gaussian in the energy difference has a width small compared with $T$,
so we adopt a procedure similar to that used in making the Sommerfeld
expansion for low temperature Fermi systems, where the
derivative of the Fermi function approaches a delta function.  In an
integral of the form,
\beq
G(x) =\int_0^\infty dy g(y)\frac{e^{-(x-y)^2/2\Delta^2}}{(2\pi
\Delta)^{1/2}},
\eeq
where $\Delta$ is a constant, and the function $g(y)$ varies slowly on the scale $\Delta$, one finds
on expanding $g$ in a Taylor series about $y=x$, that
\beq
G(x)=g(x)+\frac{\Delta^2}{2}g''(x) +\ldots \,\,.
\eeq
When this result is applied to the $q'$ integral in Eq.\
(\ref{phBE3Appendix2}) with the final Gaussian omitted, one finds
\begin{widetext}
\beq
\int_0^\infty dq  \int_0^\infty dq' \int_{-1}^1 \frac{d\cos \theta}{2}
\left(\frac{m^*s^2}{ 2\pi Tk^2}\right)^{1/2}\frac{k^2 (qq')^3(A+B\cos
\theta)^2  e^{-m^*s^2(q-q')^2/2k^2T }}{4\sinh^2(sq/2T)}\nonumber \\
\simeq J\int_0^\infty dq \frac{2q^8}{4\sinh^2(sq/2T)}
+\frac{T}{2m^*s^2}\int_0^\infty dq   \int_{-1}^1 \frac{d\cos \theta}{2}
\frac{k^2  q^3k^2(A+B\cos
\theta)^2}{2\sinh(sq/2T)}\left(\frac{\partial^2}{\partial
(q')^2}\frac{k^2q'^3}{2\sinh(sq'/2T)} \right)_{q'=q},
\label{phBE3Appendix3}
\eeq
which, when expressed in terms of the variables $x=sq/T, y=sq'/T$, and
$z=\cos\theta$, is proportional to
\beq
J\int_0^\infty dx \frac{2x^8}{4\sinh^2(x/2)}
+\frac{T}{m^*s^2}\int_0^\infty dx  \frac{x^5}{2\sinh(x/2)} \int_{-1}^1
\frac{dz}{2}(1-z) (A+Bz)^2\left(\frac{\partial^2}{\partial
y^2}\frac{\kappa^2 y^3}{2\sinh(y/2)} \right)_{y=x},
\label{phBE3Appendix4}
\eeq
where $\kappa^2=x^2+y^2-2xyz$.  The second derivative is
\beq
\frac{x^3}{ \sinh(x/2)}\left\{ 1+ (1 - z) \left[12  -\frac14 x^2- 4 x
\coth(x/2) + \frac12 x^2  \coth(x/2)^2 \right]\right\}.
\eeq
\end{widetext}
On evaluating the integrals, one finds that the effect of energy
transfer produces contributions to the integral (\ref{phBE3Appendix2})
having the form
\beq
1+  \left( 1-\frac{(25\pi^2-198)}{33} \frac{\tilde J}{J}
\right)\frac{T}{m^*s^2}.
\eeq
The numerical factor $(25\pi^2-198)/33$ is approximately 1.477 and thus
one sees that the effects of nonzero energy transfer are much less
important than those due to nonzero momentum transfer.
When both contributions to the thermal conductivity are included, one
finds on inserting Eq.\ (\ref{kph}) for the low temperature limiting
behavior,
\beq
K\simeq \frac{n_4s^2}{90\pi x_3JT}
\left(1        +
\frac{T}{m^*s^2}\left[ \frac{100\pi^2-198}{33}       \frac{\tilde J}{J}        -1\right]\right).\nonumber\\
\label{Kphrecoil}
\eeq
The coefficient of ${\tilde J}/J$ is approximately 23.90.  For $A = -1.2$
and $B = 0.70$, $J$ is 2.16, ${\tilde J}$ is $3.30$ and their ratio is
1.52.  Thus the coefficient of $T/m^*s^2$ is
35.5.   Even though $m^*s^2$ = 48.1 K, the effects of recoil are large even at temperatures well below 1 K as a consequence of the large numerical coefficient.  This coefficient
reflects the fact that  the most important contributions to the momentum transfer arise from
phonons with momenta  $\gg T/s$.  At $T$ = 0.5 K the correction is 37\%, which is
considerable.  Higher-order contributions in $T$ can be significant, and
one would expect these to reduce the deviation from the low-temperature
limiting result by an amount of relative order $(0.37)^2\sim 10\%$.  These analytic results are
in good agreement with the numerical integration of Eq.\
(\ref{phBE3Appendix2}) described above, and shown in Fig.~\ref{fig:K}.


\begin{thebibliography}{20}
\bibitem{bowleyError} Note that in the implementation
of this correction in Ref.~\cite{Bowley}, the $^3$He effective mass  is
taken to be one third of its actual value, and therefore the effects of
recoil are overestimated.
\bibitem{xto0}  One should not be alarmed by the apparent divergence as $x_3\to0$.  When $x_3$ becomes sufficiently small, the $^3$He no longer provide the mechanism for absorbing momentum from the phonons, and rather the phonon thermal conductivity becomes limited by viscous transfer of momentum from the phonons to the container walls \cite{BBPII}.
\bibitem{visc} In fact, the viscosity for flow in a circular pipe plays a role only in a thin boundary layer, which for the concentrations and temperatures under consideration is of order mm thick.
\bibitem{tempRange}
We note that the analysis here must be extended for the lower temperatures in the experiment to include scattering from the walls.  For example, the phonon mean free path is larger than the diameter of the pipe for temperatures below about 0.37~K.
\end{thebibliography}
\end{document}